\documentclass[12pt]{iopart}
\usepackage{iopams}
\usepackage{setstack}
\usepackage{graphicx}

\newcommand{\half}{\textstyle{\frac{1}{2}}}

\newcommand{\cP}{\ensuremath{\mathcal{P}}}
\newcommand{\cT}{\ensuremath{\mathcal{T}}}

\begin{document}

\title[Chaotic systems in complex phase space]
{Chaotic systems in complex phase space}

\author[Bender, Feinberg, Hook, Weir]{Carl~M~Bender${}^1$, Joshua~Feinberg${}
^2$, Daniel~W~Hook${}^3$, and David~J~Weir${}^4$}

\address{${}^1$Department of Physics, Washington University, St. Louis, MO
63130, USA\\ {\footnotesize{\tt email: cmb@wustl.edu}}}

\address{${}^2$Department of Physics, University of Haifa at Oranim, Tivon
36006, Israel and Department of Physics, Technion, Haifa 32000, Israel\\
{\footnotesize{\tt email: joshua@technion.ac.il}}}

\address{${}^3$Blackett Laboratory, Imperial College London, London SW7 2AZ,
UK\\ {\footnotesize{\tt email: d.hook@imperial.ac.uk}}}

\address{${}^4$Blackett Laboratory, Imperial College London, London SW7 2AZ,
UK\\ {\footnotesize{\tt email: david.weir03@imperial.ac.uk}}}

\date{today}

\begin{abstract}
This paper examines numerically the complex classical trajectories of the kicked rotor and the double pendulum. Both of these systems exhibit a transition to chaos, and this feature is studied in complex phase space.  Additionally, it is shown that the short-time and long-time behaviors of these two $\cP\cT$-symmetric dynamical models in complex phase space exhibit strong qualitative similarities.
\end{abstract}

\submitto{\NL}

\section{Introduction}
\label{s1}

For the past decade there has been intense activity in the field of $\cP\cT$
quantum mechanics \cite{S1,S2}. A $\cP\cT$-symmetric Hamiltonian is said to have an {\it unbroken} $\cP\cT$ symmetry if all of its eigenfunctions are also $\cP\cT$ symmetric. A Hamiltonian having an unbroken $\cP\cT$ symmetry is
physically relevant because all of its eigenvalues are real and it generates unitary time evolution. Thus, such a Hamiltonian defines a conventional quantum-mechanical theory even though it may not be Dirac Hermitian. (A linear operator is {\it Dirac Hermitian} if it remains invariant under the combined operations of matrix transposition and complex conjugation.) One can regard such non-Hermitian quantum-mechanical systems as being complex extensions of conventional quantum systems.

The interesting features of $\cP\cT$ quantum mechanics have motivated many
recent studies of $\cP\cT$ classical mechanics. In particular, solutions to
Hamilton's equations have been examined for various systems whose Hamiltonians
are $\cP\cT$ symmetric. For such systems the classical trajectories are
typically complex \cite{S3,S4,S5,S6,S7,S8,S9,S95,S10,S11,S12,S13,S14,S15}. These
trajectories can lie in many-sheeted Riemann surfaces and often have elaborate
topological structure. When the $\cP\cT$ symmetry of the quantum Hamiltonian is
not broken, the real-energy trajectories of the corresponding classical
Hamiltonian are found to be closed and periodic \cite{S8,S12}.

The purpose of this paper is to explore a new aspect of complex classical
mechanics, namely, the complex extension of chaotic behavior. Specifically, we
study two classical systems: the kicked rotor and the double pendulum. The
kicked rotor is a paradigm for studying the dynamics of chaotic systems described by time-dependent Hamiltonians \cite{R1,R1.5,R2}. The planar double pendulum is also a dynamical model whose classical motion is known to be chaotic \cite{T1}.  The Hamiltonians for both of these dynamical systems are $\cP\cT$ symmetric so long as the parameters $K$ in the Hamiltonian for the kicked rotor (\ref{e6}) and $g$ in the Hamiltonian for the double pendulum (\ref{e16}) are real. We use a variety of computational tools in order to derive the numerical results presented. The C programming language was used to implement of a fully symplectic three-stage Gauss-Legendre Runge-Kutta method for the simulation of the double pendulum, and standard functionality in Mathematica 6 was used in the study of the kicked-rotor.

This paper is organized as follows: In Sec.~\ref{s2} we define the kicked rotor
and mention briefly the transition associated with the disappearance of KAM
trajectories. In Sec.~\ref{s3} we describe the planar double pendulum and describe the analogous transition that occurs for this dynamical system. We also
reproduce the numerical work of Heyl concerning flip times. This work reveals
fractal-like structure in the plane of initial conditions \cite{T2}. Then, in
Secs.~\ref{s4} and \ref{s5} we study the short- and long-time behaviors of the
kicked rotor and the double pendulum in the complex domain, where in part our objective is to identify indicators for the transition to chaos.  We also demonstrate that these two and very different dynamical systems exhibit remarkably similar features. Section \ref{s6} contains some concluding remarks.

\section{Kicked Rotor}
\label{s2}

The Hamiltonian for the kicked rotor is \cite{R1,R1.5,R2}
\begin{equation}
H=\frac{p^2}{2I}+K\cos\,\theta\sum_{n=-\infty}^\infty\delta(t-nT),
\label{e1}
\end{equation}
where $I$ is the moment of inertia of the rotor, $p$ is its angular momentum,
and $\theta$ is the angular coordinate. As the rotor turns, it is subjected to a
periodic impulse, which is applied at times $t=0,\,\pm T,\,\pm2T,\,\ldots\,.$ The magnitude of the impulse is proportional to $K$, a constant having dimensions of angular momentum. This Hamiltonian is $\cP\cT$ symmetric because it is symmetric separately under the operation of angular reflection $\cP$,
where $\cP:\,\theta\to2\pi-\theta$ and $\cP:\,p\to-p$, and the operation of time reversal $\cT$, where $\cT:\,t\to-t$, $\cT:\,p\to-p$, and $\cT$ leaves $\theta$ invariant. (Note that angular reflection $\cP$ is not the same as spacial reflection, which maps $\theta\to\pi-\theta$.)

Hamilton's equations of motion derived from (\ref{e1}) are
\begin{eqnarray}
\frac{d\theta}{dt}=\frac{p}{I}\quad{\rm and}\quad
\frac{dp}{dt}=K\sin\theta\,\sum_{n=0}^\infty\,\delta(t-nT).
\label{e2}
\end{eqnarray}
These equations imply that the angular momentum $p$ changes discontinuously at each kick, but remains constant between kicks. As a result, the angle $\theta$
changes linearly with time $t$ between kicks and is continuous at each kick.

It is customary to denote
\begin{equation}
p_n=p(nT+0^+)\quad{\rm and}\quad\theta_n=\theta(nT+0^+).
\label{e3}
\end{equation}
Thus, $p_n$ is the angular momentum and $\theta_n$ is the angle variable
immediately after the $n$th kick. These variables satisfy the discretized version of (\ref{e2}):
\begin{eqnarray}
\theta_{n+1}=\theta_n+{T\over I}p_n\quad{\rm and}\quad
p_{n+1}=p_n+K\sin\theta_{n+1}.
\label{e4}
\end{eqnarray}

It is conventional to replace $p_n$ and $K$ by the dimensionless quantities
\begin{equation}
\frac{T}{I}p_n\to p_n\quad{\rm and}\quad\frac{T}{I}K\to K,
\label{e5}
\end{equation}
in terms of which we rewrite (\ref{e4}) in dimensionless form as
\begin{eqnarray}
\theta_{n+1}=\theta_n+p_n\quad{\rm and}\quad
p_{n+1}=p_n+K\sin\theta_{n+1}.
\label{e6}
\end{eqnarray}
This system of difference equations, which depends on a single dimensionless
real parameter $K$, is known as the {\em standard map}. It is straightforward to
show that the standard map is area-preserving in $p-\theta$ phase space. Note
that the angular variable $\theta_n$ may be taken modulo $2\pi$. It then follows
from the first equation in (\ref{e6}) that $p_n$ may also be taken modulo $2
\pi$. Thus, (\ref{e6}) maps the two dimensional torus onto itself.

The behavior of the standard map (\ref{e6}) is elaborate and has been studied
extensively \cite{R1,R1.5,R2,R3,R4,R5,R6,R7}. For small $K$ the motion in phase space is bounded and chaotic in some regions. As $K$ increases, KAM trajectories
disappear. At the critical value $K_c=0.9716\ldots$ only the KAM trajectories
with golden-mean winding number and with inverse-golden-mean winding number
remain, and the motion in phase space is still confined. For $K>K_c$ the last
bounding trajectory is destroyed and global diffusion in phase space ensues. The
critical behavior near $K_c$ has been studied intensively \cite{R4,R7}.

\begin{figure*}[t!]
\begin{center}
\includegraphics[scale=0.82, bb=0 0 600 399]{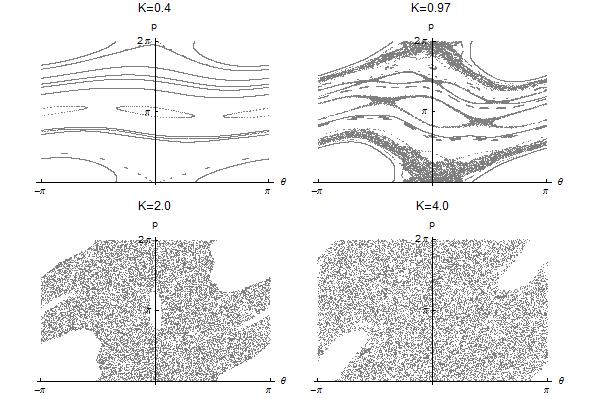}
\end{center}
\caption{Four phase-plane views of the kicked rotor. In each of the figures
we display the superposition of the discrete phase-plane trajectories $\theta_n,
p_n$ for eleven randomly chosen initial conditions. The time variable $n$
ranges from 1 to several thousand. The values of $K$ for the four plots are
0.40, 0.97, 2.0, and 4.0. The KAM surfaces separate Poincar\'e islands. Observe that as $K$ increases, the KAM surfaces gradually disappear and the trajectories diffuse into the phase plane.}
\label{f1}
\end{figure*}

Figure \ref{f1} illustrates the transition from subcritical to supercritical $K$ for the kicked rotor. In this figure we display four sets of superpositions of phase planes, each consisting of eleven randomly chosen initial conditions $\theta_0,p_0$. For each set of initial conditions we allow the time variable $n$ to range from 1 to several thousand. The values of $K$ for these four plots are 0.40, 0.97, 2.0, and 4.0.

In this paper we continue the classical dynamics described by the standard map
(\ref{e6}) into complex phase space.\footnote{The idea to study chaotic systems in complex phase space was introduced several years ago in Ref.~\cite{R8}. The motivation in these papers was to study the effects of classical chaos on semiclassical tunneling. In the instanton calculus one must deal with a complex configuration space.} Our objective here is to generalize (\ref{e6}) into complex phase space and thereby gain a better understanding of the critical behavior near $K_c$. To accomplish this we are motivated to extend the analysis of Refs.~\cite{S3,S4,S5,S6,S7,S8,S9,S95,S10,S11,S12,S13,S14,S15} to time-dependent systems. Thus, we treat $p_n$, $\theta_n$, and sometimes $K$ as complex variables, which we separate into real and imaginary parts as
\begin{equation}
p_n=r_n+is_n,\qquad\theta_n=\alpha_n+i\beta_n,\qquad K=L+iM.
\label{e7}
\end{equation}
Substituting (\ref{e7}) in (\ref{e6}), we obtain the complexified standard map
\begin{eqnarray}
\alpha_{n+1}&=&\alpha_n+r_n,\nonumber\\
\beta_{n+1}&=&\beta_n+s_n\nonumber,\\
r_{n+1}&=&r_n+L\sin\alpha_{n+1}\cosh\beta_{n+1}-M\cos\alpha_{n+1}\sinh\beta_{n+1},\nonumber\\
s_{n+1}&=&s_n+L\cos\alpha_{n+1}\sinh\beta_{n+1}+M\sin\alpha_{n+1}\cosh\beta_{n+1}.
\label{e8}
\end{eqnarray}
In Secs.~\ref{s4} and \ref{s5} we display and discuss the results of our numerical studies of (\ref{e8}).

\section{Double Pendulum}
\label{s3}

\begin{figure*}[t!]
\begin{center}
\includegraphics[scale=0.45, bb=0 0 413 511]{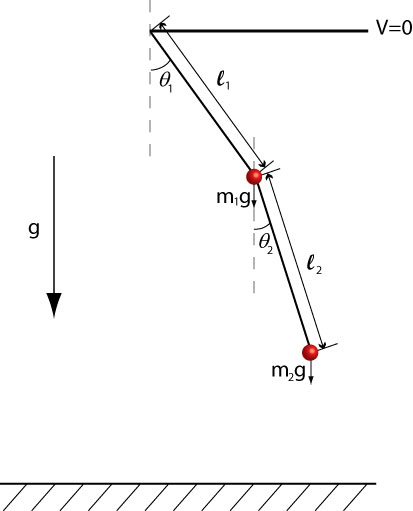}
\end{center}
\caption{Configuration of the double pendulum. The double pendulum consists of
two massless rods, each having a massive bob at the end. The second rod hangs
from the end of the first rod. The two rods are constrained to swing in a plane
and are acted on by a homogeneous gravitational field of strength $g$.}
\label{f2}
\end{figure*}

As shown in Fig.~\ref{f2}, a planar double pendulum consists of a massless rod
of length $\ell_1$ with a bob of mass $m_1$ at the lower end from which hangs a second massless rod of length $\ell_2$ with a second bob of mass $m_2$ at the lower end. This compound pendulum swings in a homogeneous gravitational field $g$, and its motion is constrained to a plane.

In this paper we take both bobs to have unit mass and both rods to have unit length. The coordinates of the bobs in terms of the angles from the vertical are
\begin{eqnarray}
x_1 &=& \sin\theta_1,\nonumber\\
y_1 &=& -\cos\theta_1,\nonumber\\
x_2 &=& \sin\theta_1+\sin\theta_2,\nonumber\\
y_2 &=& -\cos\theta_1-\cos\theta_2.
\label{e9}
\end{eqnarray}
Therefore, the potential and kinetic energies of the double pendulum are
\begin{equation}
V=-g\cos\theta_2-2g\cos\theta_1, \quad {\rm and} \quad T=\dot{\theta}_1^2+\half\dot{\theta}_2^2+\dot{
\theta}_1\dot{\theta}_2\cos(\theta_1-\theta_2).
\label{e10}
\end{equation}
From these one can form the Lagrangian $L=T-V$, and then construct the Hamiltonian for the system by a Legendre transform. We obtain
\begin{equation}
H=\frac{p_1^2+2p_2^2-2p_1p_2\cos(\theta_1-\theta_2)}{2\left[\sin^2(\theta_1-
\theta_2)+1\right]}-g\cos\theta_2-2g\cos\theta_1.
\label{e15}
\end{equation}
This Hamiltonian is $\cP\cT$-symmetric because it is symmetric separately under
the operation of angular reflection $\cP$, where $\cP:\,\theta_{1,2}\to2\pi- \theta_{1,2}$ and $\cP:\,p_{1,2}\to-p_{1,2}$, and the operation of time reversal $\cT$, where $\cT:\,t\to-t$, $\cT:\,p_{1,2}\to-p_{1,2}$, and $\cT$ leaves $\theta_{1,2}$ invariant.

Hamilton's equations are then
\begin{eqnarray}
\dot{p}_1&=&-\frac{\partial H}{\partial\theta_1}=-2g\sin\theta_1-\frac{p_1p_2
\sin(\theta_1-\theta_2)}{\sin^2(\theta_1-\theta_2)+1}\nonumber\\
&&\quad+\frac{\left[p_1^2+2p_2^2-2p_1p_2 \cos(\theta_1-\theta_2)\right]\sin\left[2(\theta_1-\theta_2)\right]}{2\left[\sin^2(\theta_1-\theta_2)+1
\right]^2},\nonumber\\
\dot{p}_2&=&-\frac{\partial H}{\partial\theta_2}=-g\sin\theta_2
+\frac{p_1p_2\sin\left(\theta_1-\theta_2\right)}{\sin^2\left(\theta_1-\theta_2
\right)+1}\nonumber\\
&&\quad-\frac{\left[p_1^2+2p_2^2-2p_1p_2\cos\left(\theta_1-\theta_2\right)\right]\sin\left[2\left(\theta_1-\theta_2
\right)\right]}{
2\left[\sin^2\left(\theta_1-\theta_2\right)+1\right]^2},\nonumber\\
\dot{\theta}_1&=&\frac{\partial H}{\partial p_1}=\frac{p_1-p_2\cos\left( \theta_1-\theta_2\right)}{\sin^2\left(\theta_1-\theta_2\right)+1},
\nonumber\\
\dot{\theta}_2&=&\frac{\partial H}{\partial p_2}=\frac{2p_2-p_1\cos\left( \theta_1-\theta_2\right)}{\sin^2\left(\theta_1-\theta_2\right)+1}.
\label{e16}
\end{eqnarray}
Note that this system conserves energy, unlike the kicked rotor whose Hamiltonian (\ref{e1}) is time-dependent.

A beautiful and convincing numerical demonstration that the motion of the double
pendulum is complicated and elaborate was given by Heyl \cite{T2}. In his work
Heyl calculates for a given initial condition the time required for either
pendulum to exhibit a {\it flip}; that is, for either $\theta_1$ or $\theta_2$ to exceed the value $\pi$. This calculation is then performed for the limited set of initial conditions for which $p_1(0)=0$ and $p_2(0)=0$, and the initial
values of $\theta_1$ and $\theta_2$ both range from 0 to $2\pi$. Each pixel in
the initial $\theta_1,\,\theta_2$ plane is then colored according to the length
of the flip time.

\begin{figure*}[t!]
\begin{center}
\includegraphics[scale=0.6, bb=0 0 401 401]{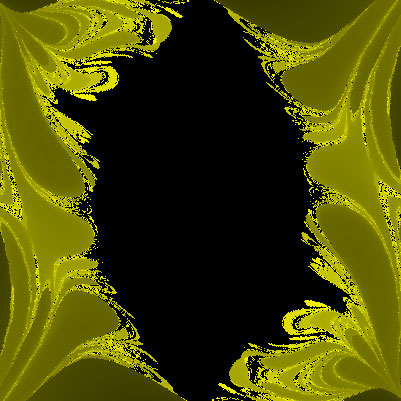}
\end{center}
\caption{Flip times for initial conditions $-\pi\leq\theta_1(0)\leq\pi$, $-\pi \leq\theta_2(0)\leq\pi$, $p_1(0)=0$, and $p_2(0)=0$. By a {\it flip} we mean that the angular position of either bob exceeds $\pi$. This figure contains a $600\times600$ grid of pixels. The color of each pixel characterizes the behavior of the double pendulum that arises from the initial condition $[\theta_1(0),\,\theta_2(0)]$. If neither pendulum flips within 100 time units, then the pixel is black (the convex-lens-shaped region in the center of the figure). If one of the pendula flips in a short time, the pixel is colored dark gray. Longer flip times are indicated by lighter shades of gray. The fractal-like structure throughout the diagram reveals the nontrivial dynamics of the double pendulum. Because of parity symmetry this figure is symmetric under the combined reflections $\theta_{1,2}\to-\theta_{1,2}$. Note that different pixels are associated with different energies and that there is an elliptical region at the center of the figure in which flips are forbidden by energy considerations.}
\label{f3}
\end{figure*}

We have applied Heyl's approach to the double pendulum in (\ref{e16}) and have used a Gauss-Legendre Runge-Kutta method, which is known to be fully symplectic \cite{U1,U2}. The results of this calculation are given in Fig.~\ref{f3}. This figure is composed of a $600\times600$ grid of pixels,
where each pixel represents the initial condition $[\theta_1(0),\,\theta_2(0)]$.
If neither pendulum flips within 100 time units, then the pixel is assigned the
color black (the convex-lens-shaped region in the center of the figure). If
either pendulum flips in a short time, the pixel is colored dark gray, with longer flip times being indicated by lighter shades of gray. Notice the fractal-like structure throughout the diagram. The appearance of this complicated structure demonstrates that even though the double pendulum has only two degrees of freedom, it exhibits rich and nontrivial dynamics.

\begin{figure*}[t!]
\begin{center}
\includegraphics[scale=0.75, bb=0 0 600 420]{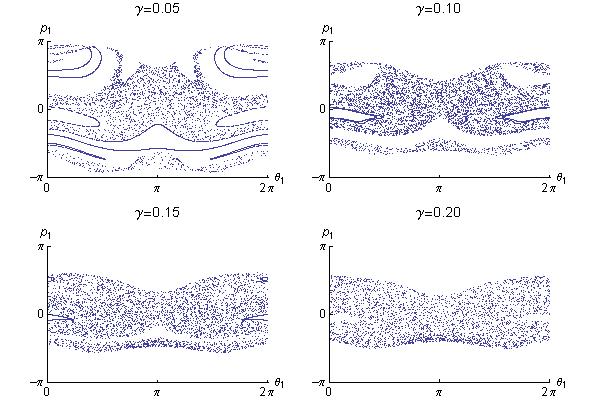}
\end{center}
\caption{Poincar\'e plots for the double pendulum for four values of $\gamma$. Below the critical value, which is near $\gamma=0.1$, the final KAM surface can still be seen in the second plot, but this surface evaporates as $\gamma$ increases past the critical value and the points in the plot spread. This plot is analogous to that in Fig.~\ref{f1} for the kicked rotor.}
\label{fx2}
\end{figure*}

In analogy with the kicked rotor, there is a transition in the behavior of the double pendulum in which KAM surfaces disappear as a dimensionless parameter increases beyond a critical value. This parameter, which measures the strength of the gravitational field relative to the total energy, is defined as \cite{T1}
\begin{equation}
\gamma\equiv m_1g\ell_1/E.
\label{e17}
\end{equation}
The transition occurs near $\gamma=0.1$ and Ref.~\cite{T1} shows that at the transition the last surviving KAM surface is the one with winding number being equal to the golden mean. In Fig.~\ref{fx2} we plot the Poincar\'e sections generated from 25 randomly chosen initial conditions for four different values of $\gamma$. The plot displays points in the $\theta_1,p_1$ plane when $\theta_2=0$ and simultaneously $p_2>0$. A KAM surface is visible when $\gamma=0.05$, which is below the critical value. At $\gamma=0.1$, which is near the critical value, the KAM surface disappears. For the other two values of $\gamma$, which are significantly greater than the critical value, the distribution of points in the plot becomes diffuse in a manner analogous to the behavior displayed in Fig.~\ref{f1} for the kicked rotor in the $K>K_c$ regime.

\section{Short-time behavior}
\label{s4}

Having reviewed some of the well-known properties of the kicked rotor and the
double pendulum, we now proceed to examine the behavior of the solutions to the
kicked-rotor and double-pendulum equations of motion in the complex domain. To do so, we do not change the form of the equations of motion, but rather we take complex initial conditions and in some cases we allow the parameter $K$ in (\ref{e6}) for the kicked rotor and the parameter $g$ in (\ref{e16}) for the double pendulum to take on complex values.

\begin{figure*}[t!]
\begin{center}
\includegraphics[scale=0.75, bb=0 0 500 629]{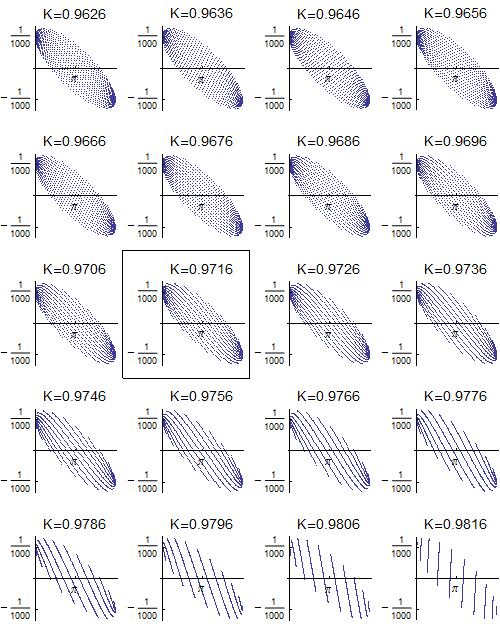}
\end{center}
\caption{Behavior of the solution to the kicked rotor in the complex-$\theta$
plane for a range of real values of $K$ near $K_c=0.9716\ldots\,.$ We take as an
initial condition $p_0=0$ and $\theta_0=1+0.0001i$ and allow the system to
evolve for $n=1000$ time steps. For each value of $n$ we plot the complex value
of $\theta_n$ as a point in the complex-$\theta$ plane. The plot corresponding
to the critical value is highlighted. Note that there is a qualitative change in the nature of these complex plots as $K$ passes through its critical value. Specifically, the points making up the plot become less uniform and more stratified. These changes in behavior are easier to observe than those in Fig.~\ref{f1}.}
\label{f4}
\end{figure*}

In this section we investigate the behavior of these dynamical systems for short
times; that is, for up to 1000 time steps. For the kicked rotor, let us see what
happens if we take the initial momentum to be real, $p_0=0$, but take the
initial angle to have a small imaginary component, $\theta_0=1+0.0001i$. In
Fig.~\ref{f4} we plot the points $\theta_n$ in the complex-$\theta$ plane for $n=0,\,1,\,2,\,\ldots,\,1000$ for a range of {\it real} values of $K$ around the critical point $K_c=0.9716\ldots\,.$ While it is difficult to see the subtle change from subcritical to supercritical behavior in real plots like those in Fig.~\ref{f1}, a qualitative change in the complex behavior is quite evident in Fig.~\ref{f4} \cite{TALK}. Below $K_c$ the points tend to occupy a two-dimensional region in the complex plane with some fine structure in it, but as $K$ increases above $K_c$, points tend to coalesce along well separated one-dimensional curves.

\begin{figure*}[t!]
\begin{center}
\includegraphics[scale=0.67, bb=0 0 600 737]{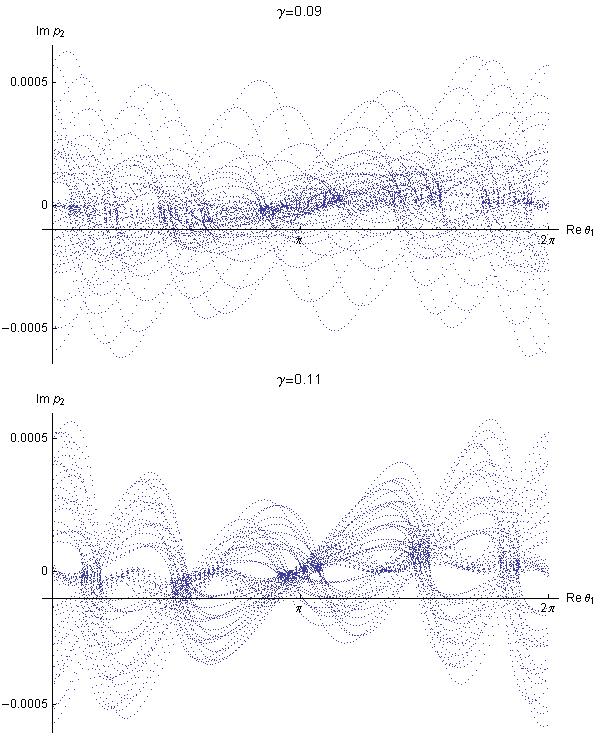}
\end{center}
\caption{Trajectory for the double pendulum that results from the slightly complex initial conditions $\theta_1=3$, $\theta_2=3+0.0001i$, $p_1=p_2=2$. The parameter $\gamma$ is varied by changing the values of $g$ and $l_1$ in relation (\ref{e17}) and fixing $m_1=E=1$.  This plot displays the trajectory in the ${\rm Re}\, \theta_1, \,{\rm Im}\,p_2$ plane as a function of $t$ for $0\leq t\leq1000$. The trajectory is confined to narrow bands when $\gamma$ is above the critical value and is more diffuse when it is below the critical value. In this sense, this figure has features that are similar to those in Fig.~\ref{f4} for the kicked rotor.}
\label{fx1}
\end{figure*}

In analogy with Fig.~\ref{f4}, we plot in Fig.~\ref{fx1} a trajectory for the double pendulum in the ${\rm Re}\, \theta_1,\,{\rm Im}\,p_2$ plane as a function of $t$ for $0\leq t\leq224$. We take two values of $\gamma$, one subcritical and one supercritical and use the slightly complex initial conditions $\theta_1=3.1$, $\theta_2=3.1+0.0001i$, $p_1=1.283$, and $p_2=1.283$ in which the numbers are chosen at random. Observe that when $\gamma$ is above the critical value the trajectory appears to be confined to distinct narrow bands, somewhat similar to the stratified structures that occur for $K>K_c$ in Fig.~\ref{f4} for the kicked rotor. However, when $\gamma$ is below the critical value, the trajectory spreads and is similar to the more diffuse behavior in Fig.~\ref{f4} when $K<K_c$.  It would be interesting to understand this stratification analyticallty and also to understand its relation to KAM theory.

An even more dramatic way to observe the transition from the subcritical to the
supercritical regions of the kicked rotor is to construct plots like that in
Fig.~\ref{f3} for the double pendulum. We take as an initial condition $p_0=0$ and take a range of complex initial values for $\theta_0$: $-\pi\leq{\rm Re}\, \theta_0\leq\pi$ and $-\pi\leq{\rm Im}\,\theta_0\leq\pi$. For each initial condition we then allow the kicked rotor to evolve up to a maximum of 400 time steps and determine the time step at which the real part of $p_n$ becomes infinite, if it does actually become infinite. (Here, by {\it infinite} we mean that the numerical value of ${\rm Re}\,p_n$ exceeds $10^{308}$, which is the largest number that may be represented in double precision arithmetic.) We then perform this calculation for each pixel on a $628\times628$ grid representing the complex-$\theta_0$ plane. We assign a color to each pixel corresponding to the time at which $p_n$ becomes infinite: White indicates that $p_n$ does not become infinite within 400 time steps, and darker shades indicate that $p_n$ becomes infinite after shorter times.

\begin{figure*}[t!]
\begin{center}
\includegraphics[scale=0.74, bb=0 0 600 600]{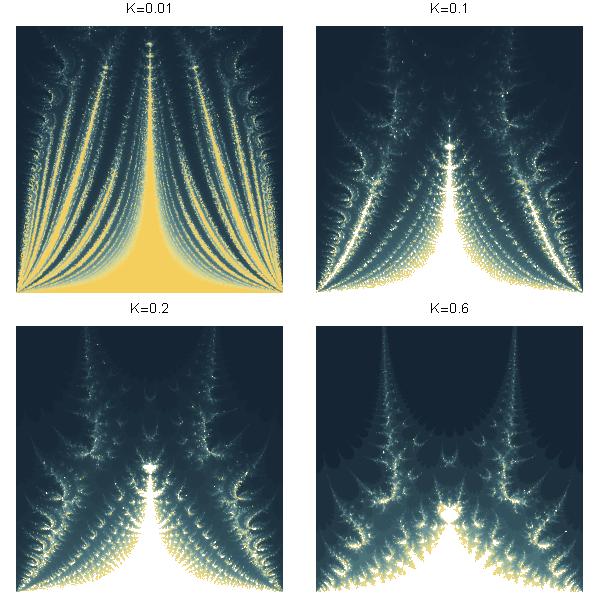}
\end{center}
\caption{Behavior of the solution to the kicked rotor in the complex-$\theta$
plane for four real values of $K$ and a range of complex initial conditions: $p_0=0$, $0\leq{\rm Re}\,\theta_0\leq2\pi$, and $0\leq{\rm Im}\,\theta_0\leq2 \pi$. We allow the system to evolve for at most $n=400$ time steps. We assign a color to each value of $\theta_0$ according to the time at which ${\rm Re}\, p_n$ becomes infinite (if it does so). The graph of initial values of $\theta$ clearly has fractal structure. Distinctive changes occur in the nature of these complex plots as $K$ increases.}
\label{f5}
\end{figure*}

\begin{figure*}[t!]
\begin{center}
\includegraphics[scale=0.74, bb=0 0 600 600]{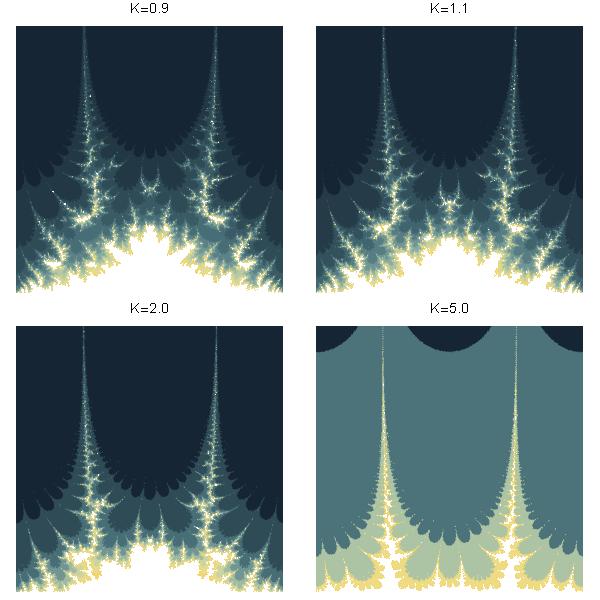}
\end{center}
\caption{Same as in Fig.~\ref{f5} with four higher values of $K$. There are many qualitative changes that occur as $K$ increases. For example, the dendritic landscape that occurs for smaller values of $K$ becomes smoother and more rounded as $K$ increases. A subtle but important change is that the white regions, the regions in which ${\rm Re}\,p_n$ does not diverge for $n\leq400$, become connected when $K$ exceeds $K_c$.}
\label{f6}
\end{figure*}

In Fig.~\ref{f5} we display the results of this calculation for $K=0.01,\,0.1,\,0.2$, and $0.6$, and in Fig.~\ref{f6} we display the results for this calculation for $K=0.9,\,1.1,\,2.0$, and $5.0$. Note that all these figures exhibit a complicated dendritic and fractal-like structure. A number of qualitative changes occur as $K$ increases past its critical value. One obvious change is that the dendritic landscape becomes smoother and more rounded as $K$ increases. A less obvious change is that the regions in which ${\rm Re}\,p_n$ does not diverge for $n\leq400$ become connected when $K$ exceeds $K_c$.

\begin{figure*}[t!]
\begin{center}
\includegraphics[scale=0.75, bb=0 0 360 380]{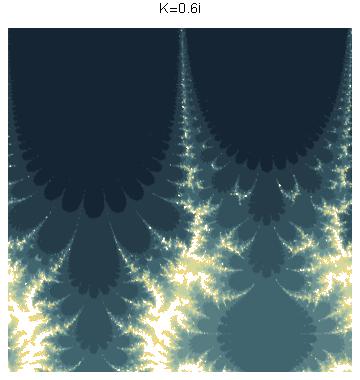}
\end{center}
\caption{Same as in Figs.~\ref{f5} and \ref{f6} except that $K=0.6i$. Observe that while many features of the fractal structure in Figs.~\ref{f5} and \ref{f6} are preserved, they are distorted and the left-right symmetry is destroyed.}
\label{f7}
\end{figure*}

Instead of constraining $K$ to be real, we can, of course, take $K$ to be complex. In Fig.~\ref{f7} we take $K=0.6i$. Note that the fractal structure in Figs.~\ref{f5} and \ref{f6} is preserved, but it is distorted and loses its left-right symmetry.

\begin{figure*}[t!]
\begin{center}
\includegraphics[scale=0.74,bb=0 0 600 600]{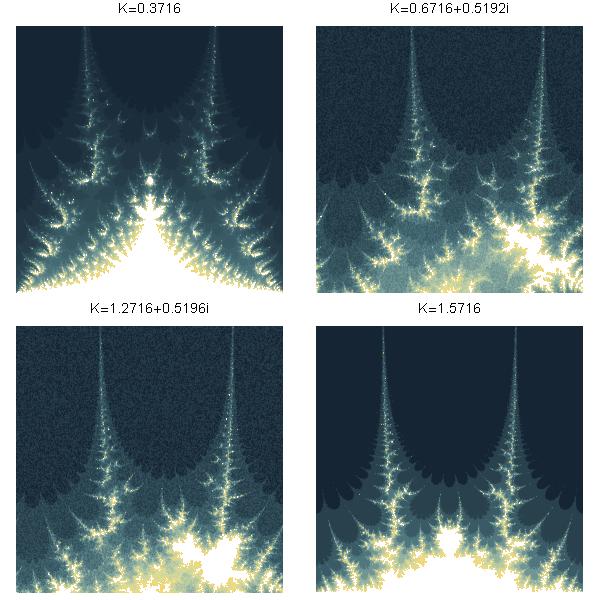}
\end{center}
\caption{Same as in Fig.~\ref{f5} except that the pictures making up this figure are constructed from values of $K$ lying on a semicircle of radius $0.6$ and centered at the critical value $K_c=0.9716\ldots\,.$ In these figures we observe fractal structures similar to those in Figs.~\ref{f5}-\ref{f7}, but slightly distorted. An important difference between this figure and Figs.~\ref{f5}-\ref{f7} is that there is mottling (speckling) in those graphs where $K$ is complex.}
\label{f8}
\end{figure*}

\begin{figure*}[t!]
\begin{center}
\includegraphics[scale=0.75, bb=0 0 450 480]{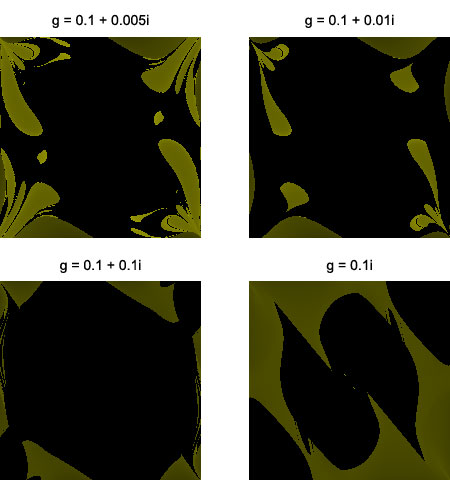}
\end{center}
\caption{Short-time behavior of the double pendulum with complex $g$. Same as Fig.~\ref{f3} but with $g=0.1+0.005i,\,0.1+0.01i,\,0.1+0.1i,\,0.1i$. Note that as ${\rm Im}\,g$ increases, the fractal-like structure that we see in Fig.~\ref{f3} moves outward and towards the corners, and the boundaries between differently colored regions become smooth. In contrast with Fig.~\ref{f3}, there are no energetically-forbidden-flip regions because there exist complex pathways from any pixel to a flipped configuration.}
\label{f9}
\end{figure*}

Rather than requiring that $K$ pass its critical value on the real-$K$ axis, it is possible to go from a subcritical real value to a supercritical real value via a path in the complex-$K$ plane, as in Fig.~\ref{f8}. The pictures making up this figure are constructed from values of $K$ that lie on a semicircle of radius $0.6$ and are centered at the critical value $K_c=0.9716\ldots\,.$ In these figures we observe fractal structures like those in Figs.~\ref{f5}-\ref{f7}, but they are slightly distorted. However, there is a significant difference in that there is mottling (replacement of large patches of solid shading by a speckled pattern) in the graphs where $K$ is complex; this mottling is absent when $K$ is pure real or pure imaginary.  (Note that when $K$ is complex, $\cP\cT$ symmetry is broken if the $\mathcal{T}$ operator is antilinear, that is, it changes the sign of $i$.\footnote{When $K$ and $g$ become pure imaginary the system becomes invariant under {\it combined} $\cP\cT$ reflection. However, now $\cP$ is a spatial reflection, $\cP:$ $\theta\to\theta+\pi$, so that both $\cos\theta$ and $\sin\theta$, and thus the cartesian coordinates, change sign. The sign of the angular momentum now remains unchanged under parity reflection. This explains the symmetry of the plots when $K$ and $g$ are pure imaginary (see Fig.~\ref{f7} and the lower-right plot in Fig.~\ref{f9} respectively).  This change of symmetry of the system as its couplings vary in complex parameter space is not unusual. For example, at a generic point in coupling space for the three-dimensional anisotropic harmonic oscillator, the only symmetry is parity. However, when any two couplings coincide and are different from the third, the reflection symmetry is enhanced and becomes a  continuous symmetry, namely, an $O(2)$ symmetry around the third axis. (There remains parity-time reflection symmetry in the third direction.)  When all three couplings coincide, the symmetry is enhanced further and becomes a full $O(3)$ symmetry.})

We have performed a closely related study for the double pendulum: We allow $g$ to be complex and repeat the numerical analysis of the short-time behavior that we used to produce Fig.~\ref{f3}. We find that as the imaginary part of $g$ increases, the fractal-like structure that we see in Fig.~\ref{f3} gradually moves outward towards the corners of the figure. Correspondingly, the boundaries between different colored regions become smoother. To demonstrate this effect, we choose four different complex values, $g=0.1+0.005i,\,0.1+0.01i,\,0.1+0.1i,\,0.1i$, and plot the results in Fig.~\ref{f9}.

\section{Long-time behavior}
\label{s5}
\begin{figure*}[t!]
\begin{center}
\includegraphics[scale=0.6, bb=0 0 600 939]{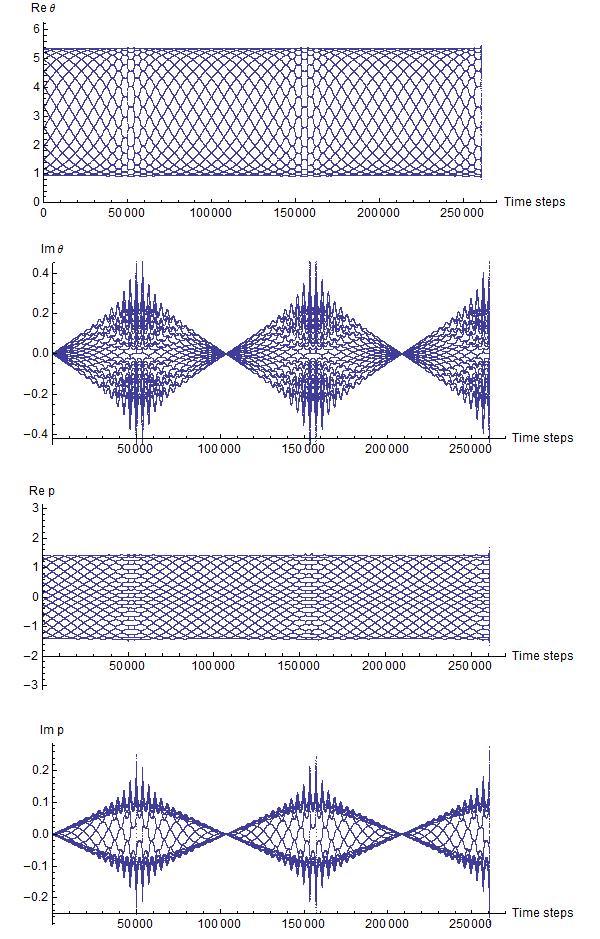}
\end{center}
\caption{Long-time behavior of the kicked rotor with initial conditions $p_0=0$ and $\theta_0=1+10^{-5}i$ and $K=0.6$. Note that while the real parts of $p_n$ and $\theta_n$ oscillate between almost constant boundaries, the imaginary parts of $p_n$ and $\theta_n$ exhibit a synchronized ringing behavior whose period is of order $1/\left({\rm Im}\,\theta_0\right)$. In addition to the long-time ringing there is a short-time oscillation that becomes most pronounced when the amplitude of the ringing is at a maximum. After about $2\half$ rings the solution abruptly diverges and ceases to exist.}
\label{f10}
\end{figure*}

In this section we study the long-time behavior of the kicked rotor and the double pendulum in complex phase space and we find that they share many qualitative features in this regime as well. By {\it long-time} we mean roughly $10^4$ to $10^5$ time steps or time units rather than the $10^3$ time steps taken in Sec.~\ref{s4}. We find that the solutions to the dynamical equations for these systems exhibit characteristic behaviors at different time scales. On a long-time scale, which is determined by the imaginary part of the initial value of an angle, the solutions tend to {\it ring}; that is, the envelope of the solution grows and decays to zero with gradually changing periods. On a short-time scale the solution exhibits a distinct and clearly identifiable rapid oscillation, as we can see in Fig.~\ref{f10}.

We have chosen here to use the language of multiple-scale perturbation theory \cite{BO} to describe this oscillatory behavior. However, for both the kicked rotor and the double pendulum the unperturbed equations are not linear, and thus the usual techniques of multiple-scale perturbation theory cannot be applied directly in these cases.

\begin{figure*}[t!]
\begin{center}
\includegraphics[scale=0.66, bb=0 0 600 332]{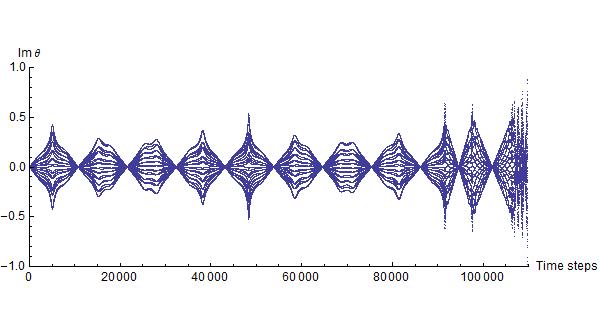}
\end{center}
\caption{Same as in Fig.~\ref{f10} except that the imaginary part of $\theta_0$ is ten times smaller: $\theta_0=1+10^{-4}i$ and only the imaginary part of $\theta_n$ is displayed. In this figure the period of the ringing is ten times shorter and there are about eleven rings before the solution destabilizes and ceases to exist.}
\label{f11}
\end{figure*}

Let us first examine the kicked rotor. As an initial condition we choose $p_0= 0$ and $\theta_0=1+i\epsilon$. In Fig.~\ref{f10} we take $K=0.6$ and $\epsilon=10^{-5}$ and we plot ${\rm Re}\,\theta_n$, ${\rm Im}\,\theta_n$, ${\rm Re}\, p_n$, and ${\rm Im}\,p_n$ for $0\leq n\lesssim260\,000$. Note that while ${\rm Re}\,\theta_n$ and ${\rm Re}\,p_n$ oscillate within almost constant boundaries, ${\rm Im}\,\theta_n$ and ${\rm Im}\,p_n$ appear to ring with a period of order $1/\epsilon$. To verify this dependence on $\epsilon$ we take $\epsilon=10^{-4}$, which is ten times larger, and we do not change the other initial conditions or the value of $K$. The result for ${\rm Im}\,\theta_n$ is given in Fig.~\ref{f11}, where we see that the period of the ringing is roughly ten times shorter than the period in Fig.~\ref{f10}. In both of these figures the ringing eventually comes to an abrupt end, at which point the iteration diverges and the amplitude of oscillation becomes infinite; this happens after about $2\half$ rings in Fig.~\ref{f10} and after about eleven rings in Fig.~\ref{f11}.

\begin{figure*}[t!]
\begin{center}
\includegraphics[scale=0.6, bb=0 0 600 939]{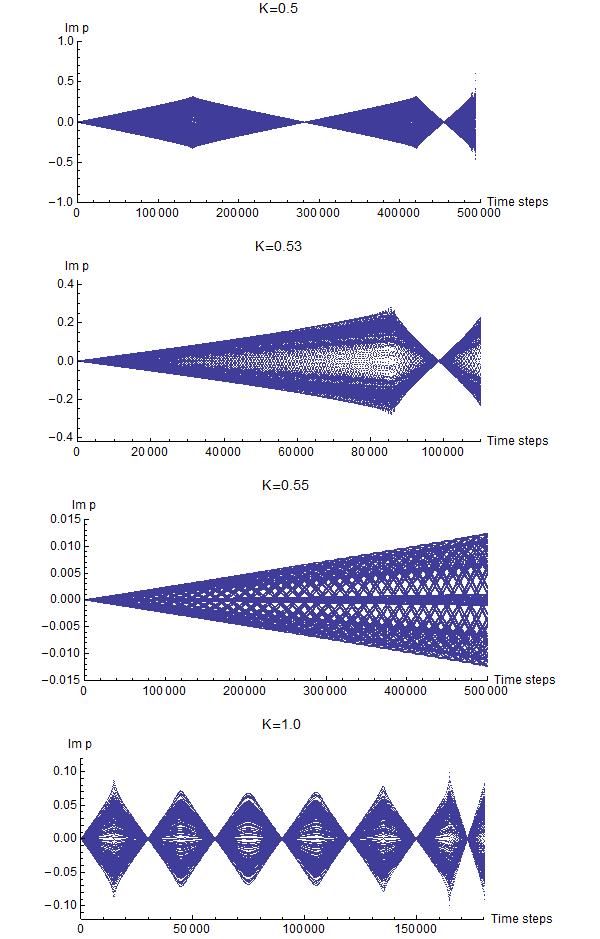}
\end{center}
\caption{Long-time behavior of ${\rm Im}\,p_n$ for the kicked rotor for four different values of $K$. The initial conditions are the same as in Fig.~\ref{f10}. Observe that the period of ringing is quite sensitive to the
value of $K$.}
\label{f12}
\end{figure*}

While the inverse of the imaginary part of $\theta_0$ appears to set the scale of the ringing period, we have found that the length of the ringing period is also sensitive to the value of $K$. In Fig.~\ref{f12} we plot ${\rm Im}\,p_n$ for four values of $K$: 0.5, 0.53, 0.55, and 1.0. For each of these values of $K$ we plot ${\rm Im}\,p_n$ until it diverges. The initial conditions for each graph in Fig.~\ref{f12} are the same as those in Fig.~\ref{f10}.

\begin{figure*}[t!]
\begin{center}
\includegraphics[scale=0.66, bb=0 0 600 332]{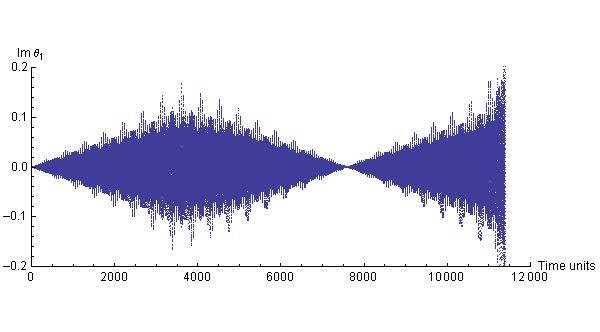}
\end{center}
\caption{Long-time behavior of the double pendulum with $g=1$. The initial conditions used for this plot are $p_1(0)=p_2(0)=0$, $\theta_1(0)=1$, and $\theta_2(0)=10^{-4}i$. Like the kicked rotor, the imaginary part of an angle exhibits a ringing behavior whose characteristic period is of order $1/\left( {\rm Im}\,\theta_2(0)\right)$. The plot terminates when the solution to the equations of motion abruptly diverges. Similar to the behavior of the kicked rotor, there is also a short-time oscillation, but unlike the kicked rotor, the positive and negative peaks are out of phase with one another.}
\label{f13}
\end{figure*}

\begin{figure*}[t!]
\begin{center}
\includegraphics[scale=0.66, bb=0 0 600 332]{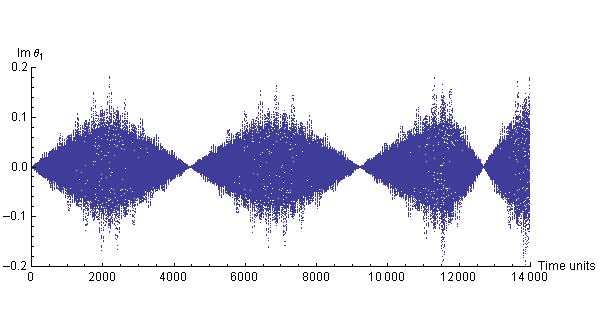}
\end{center}
\caption{Same as Fig.~\ref{f13} but with $\theta_2(0)=2\times10^{-4}i$. Note that doubling the imaginary part of $\theta_2(0)$ has the effect of roughly halving the ringing period.}
\label{f14}
\end{figure*}

The double pendulum exhibits a long-time ringing behavior that almost exactly parallels that of the kicked rotor. We plot the long-time behavior of ${\rm Im}\,\theta_1$ for $g=1$ and initial conditions $p_1(0)=p_2(0)=0$, $\theta_1(0) =1$, and $\theta_2(0)=10^{-4}i$ in Fig.~\ref{f13} and for $\theta_2(0)=2\times 10^{-4}i$ in Fig.~\ref{f14}. Note that, like the kicked rotor, the long-time-scale ringing periods are determined by the imaginary part of the initial value of an angle; here, the period is proportional to $1/\left[{\rm Im}\,\theta_2(0)\right]$.

\section{Concluding remarks}
\label{s6}
Apart from making the obvious remark that the two nonlinear systems studied in this paper exhibit very similar short-time and long-time dynamical behaviors,
this work indicates that studying the dynamics of classical chaotic systems in {\it complex} phase space may help us to understand the onset of chaos. For example, for the case of the kicked rotor, we observe in Fig.~\ref{f4} a change
in the complex behavior as $K$ increases past $K_c$. Of course, the results reported here are empirical, but they clearly underscore the need for a deeper analytical understanding of these models. For example, an important unanswered question is, what is the analog of the KAM theorem in complex phase space?

Finally, we remark that the kicked rotor is one of the rare time-dependent systems whose quantum dynamics may be studied in some detail. Indeed, the kicked rotor is a paradigm for studying quantum chaos. It might be particularly useful to explore the $\cP\cT$-deformed analog of the work of Fishman et al. \cite{R2} because (i) this would be a nontrivial extension of $\cP\cT$ quantum mechanics to time-dependent systems, and (ii) it may provide a way to define and understand $\cP\cT$-symmetric quantum chaos.

\vspace{0.5cm}
\footnotesize
\noindent
We thank S.~Fishman and F.~Leyvraz for several informative discussions and I.~Guarnery for bringing Ref.~\cite{R8} to our attention. CMB is supported by a grant from the U.S.~Department of Energy. JF thanks the KITP at UC Santa Barbara for its kind hospitality while this paper was completed. His research at the KITP was supported in part by the National Science Foundation under Grant No. PHY05-51164. DWH is supported by Symplectic Ltd. DJW thanks the Imperial College High Performance Computing Service, URL:
http://www.imperial.ac.uk/ict/services/teachingandresearchservices/highperormancecomputing.

\end{document}